\documentclass{vgtc}                          

\ifpdf
  \pdfoutput=1\relax                   
  \usepackage{graphicx}                
  \DeclareGraphicsExtensions{.pdf,.png,.jpg,.jpeg} 
\else
  \usepackage{graphicx}                
  \DeclareGraphicsExtensions{.eps}     
\fi%

\graphicspath{{figures/}{pictures/}{images/}{./}} 

\usepackage{microtype}                 
\PassOptionsToPackage{warn}{textcomp}  
\usepackage{textcomp}                  
\usepackage{mathptmx}                  
\usepackage{times}                     
\usepackage{cite}                      
\usepackage{tabu}                      
\usepackage{booktabs}                  
\usepackage[usenames, dvipsnames]{color}
\usepackage{enumitem}

\usepackage{fixltx2e}
\usepackage[skip=3pt]{caption}
\usepackage{wrapfig}

\usepackage{verbatim}

\usepackage[utf8]{inputenc}

\usepackage[font={scriptsize,sf}]{caption}

\onlineid{1011}

\vgtccategory{Research}

\vgtcinsertpkg

\title{3De Interactive Lenses for Visualization in Virtual Environments}

\author{Roberta C. R. Mota 
\thanks{\textit{The authors are with the University of Calgary, Canada.
E-mails: \{roberta.cabralmota, acarocha, jd.silva, ualim, ehud\}@ucalgary.ca}}
\and Allan Rocha %
\and Julio Daniel Silva %
\and Usman Alim %
\and Ehud Sharlin}

\teaser{
\centering 
  \includegraphics[width=\textwidth]{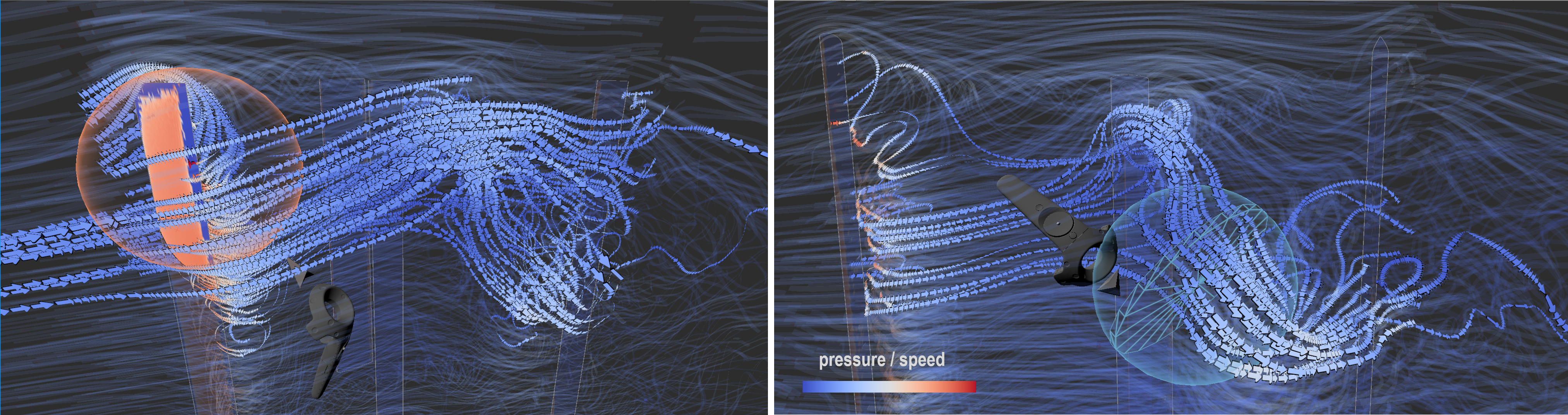}
  \caption{\textit{3De lens} inspecting wind turbine aerodynamics. (\textit{left}) The lens depicts strong-pressure gradient on the leftmost wind turbine blade; it also selects the air flowing around this blade and advancing to the rightmost one. (\textit{right}) By setting its orientation, the lens selects a disorderly bundle of flow lines passing through the central blade.}
  \label{fig:airFlow}
}

\abstract{%
We present \textit{3De lens}, a technique for focus+context visualization of multi-geometry data. It fuses two categories of lenses (3D and Decal) to become a versatile lens for seamlessly working on multiple geometric representations that commonly coexist in 3D visualizations. In addition, we incorporate our lens into virtual reality as it enables a natural style of direct spatial manipulation for exploratory 3D data analysis. To demonstrate its potential use, we discuss two domain examples in which our lens technique creates customized visualizations of both surfaces and streamlines.
} 

\CCScatlist{
    \CCScatTwelve{Human-centered computing}{Human computer interaction}{Interaction paradigms}{Virtual Reality};
  \CCScatTwelve{Human-centered computing}{Visu\-al\-iza\-tion}{Visualization Application Domains}{Scientific Visualization}
}

\begin{document}


\firstsection{Introduction}

\maketitle


In 3D visualizations, data is often represented as a combination of geometric representations (e.g., points, streamlines, surfaces), with associated attributes. To present data across multiple representations brings on the need for investigating the (inter-relations of) different data types to obtain an integrated understanding of the phenomenon. Yet, the complexity associated with multi-geometry data inevitably causes significant visual clutter and occlusion; this impedes users to properly analyze the data space.

To address the challenges of clutter and occlusion, the visualization community has embraced the idea of interactive lenses: focus+context tools that define focus areas of the visual data representation; focus is shown in details while remaining regions convey context. More precisely, a lens is an \textit{interactively parametrizable spatial selection} according to which a base visualization is altered \cite{Tominski:2017}. Key geometric parameters of lenses include shape and transformations (position, orientation, and scale). Shape is the most prominent property and commonly drives the spatial selection; position and scale attune the lens to different parts of the data; orientation is often an overlooked parameter but can be useful to align the lens in accordance with the underlying visualization, or even to fine-tune the spatial selection based on a directional parameter. We here classify existing lenses for 3D data according to their geometric parameters as 2D, 2.5D, 3D, and Decal.

Most techniques are so-called 2D lenses: polygonal shapes (e.g., circular) placed and manipulated in screen space. Their 2D nature constrains the lens transformations to the view plane, and leads to two major drawbacks: an inability to carry out 3D selection, and a lack of spatial correlation between the 2D position and orientation of the lens and the underlying 3D visualization. 2.5D lenses consist of a 2D lens embedded in 3D space. 2.5D lens transformations are extended to 3D; this alleviates the issue of spatial correlation. It introduces, however, increased interaction effort to place and align the lens in the 3D data set. In addition, the flat nature of the lens constrains the spatial selection to only a slice of the 3D data set.

An extension of the 2.5D lens is 3D lens, which consists of a 3D volume (e.g., sphere) onto which lens operators are applied. 3D lenses are usually used for spatial selection of volumetric data; for surface data, however, such lenses do not closely align with the underlying geometry due to their rigid shape. This can lead to perceptual issues such as the lens itself occluding parts of the surface (especially those of intricate geometry). Moreover, the manipulation of a 3D lens suffers from interaction effort like a 2.5D lens.

Another lens category, called Decal, supports selection of surface areas of interest. A recently proposed decal-lens~\cite{decalLenses} is built from the intersection of a sphere with a surface. The sphere itself does not provide the area where the lens effect takes place; rather, the lens refers to the part of the surface that lies inside the sphere. This leads to a lens-region that follows the surface geometry, like a decal. This lens allows for 3D manipulation due to its intrinsic volumetric shape; but the lens effect manifests solely on surfaces.

Given the abovementioned, we note two key remarks: 3D and decal-lenses may share common parameters (e.g., a spherical shape), and are somewhat complementary in terms of data types they best operate on (non-surface and surface, respectively). Beyond that, to date, lens techniques are commonly designed for a specific data type of the underlying visualization \cite{Tominski:2017}. In this regard, multi-geometry 3D visualizations can use separate lenses for different data types; yet, this may increase user's effort to manipulate each lens individually and lead lenses in close proximity to occlude each other.

This work thus contributes \textit{3De lens}, a technique that combines 3D and Decal lenses into one coherent entity to act on multi-geometry data. Lens manipulations and parameter adjustments become centralized, mutual occlusion no longer holds, and the lens seamlessly works on distinct geometric representations that commonly coexist in 3D visualizations. We illustrate the applicability of our lens through two domain examples to specifically support the visual exploration of surfaces and streamlines.


Moreover, in terms of interaction modality, most lens techniques are designed for classic desktop workstations with mouse and keyboard. In 3D visualizations, particularly, these systems face the fundamentally difficult problem of manipulating the lens in 3D within a 2D interaction space. For instance, one may need to place lenses to cut and reveal data values in the cavities of the aorta region of the heart;  to seed particles and  interrogate intricate blood flow patterns on an anatomical model of the left ventricle; etc. Nonetheless, to adjust the lens in relation to the underlying data using standard keyboard and mouse becomes demanding---in this regard, a recent survey by Tominski \textit{et al.} \cite{Tominski:2017} emphasizes the need for adopting lenses in novel visualization environments while citing first promising research.
This work incorporates the 3De lens technique in consumer-grade virtual reality (VR) technology to enable direct lens interaction using an associative 3D interaction space; this may lead to more natural input mapping with less physical and mental strain.

Our main contributions are:
\begin{itemize}[leftmargin=*]
\setlength{\itemsep}{0pt}
\setlength{\parskip}{0.0pt}

\item The concept of \textit{3De lens} that fuses 3D and Decal lenses in order to seamlessly operate on multi-geometry 3D visualizations.



\item The visual and interaction design and implementation of our lens technique using VR technology. We discuss both the design process and technical challenges to demonstrate that, compared to widely-used desktop systems, VR presents a promising medium to ease interaction effort during lens-based 3D data exploration.

\end{itemize}

\section{Related Work}
In this section, we review work that proposes lenses for 3D data according to the geometry and the interaction modality used as medium for interactive lens techniques. For a more extensive review of lenses, we refer the reader to the survey by Tominski \textit{et al.} \cite{Tominski:2017}.

\paragraph{\textit{Lens geometry.}}  2D screen-space lenses have been useful for 3D data exploration \cite{Tong:2016, Wang:2005}. For example, as the correlation between hemodynamic attributes is necessary to aneurysm analysis, Gasteiger \textit{et al.} \cite{Gasteiger:2011} proposed \textit{FlowLens} for focus+context visualizations of pairs of attributes like wall shear stress and inflow jet. 2.5D lenses have been mostly used to navigate and to perform local (focus) operations on slices of volume data sets \cite{Coffey:2012, Niebling:2017}. Van Pelt \textit{et al.} \cite{VanPelt:2016} proposed cross-section techniques to explore 4D PC-MRI blood-flow data. After the selection of probe cross-sections on vessel models, physicians can employ techniques to locally capture blood-flow dynamics such as exploded planar reformats, streamline or pathline seeding templates, and flow-rate arrow-trails. 3D lenses were first introduced by Viega \textit{et al.} \cite{Viega:1996} to limit the lens effect to a finite subvolume of interest; this lens concept received a number of extensions for different data types. Fuhrman and Gr\"{o}ller \cite{Fuhrmann:1998}, for instance, are among the first to propose 3D lenses for 3D flow visualization. They use a volumetric box to implicitly or explictly define a spatial selection in the flow to depict higher-resolution streamlines. Rocha \textit{et al.}~\cite{decalLenses} recently proposed decal-lenses to facilitate on-demand multivariate data exploration on surfaces; to the best of our knowledge, it is the only work fitting into Decal lens category. \textit{3De lens} does not fit into the previous individual classification; rather, it merges lenses of distinct categories into a single coherent entity.

\paragraph{\textit{Lens interaction.}}

Existing lens techniques are usually designed for classic desktop visualization settings with mouse and keyboard interaction. It is used in many of the previously examined research works, for instance in \cite{Gasteiger:2011, Wang:2005, VanPelt:2016}. There has also been work towards interactive lenses for touch-enabled devices such as tabletops and high-resolution large displays  \cite{Coffey:2012, Somanath:2014, Tong:2016}. Tangible objects were also coined as physical lenses, typically used jointly with tabletops to facilitate tracking strategies and use the space surrounding the tangibles as context visualization \cite{Spindler:2009, Steimle:2013}. One limitation that both touch- and mouse-based interaction face is the problem of 3D data manipulation within a 2D interaction space. Tangible interaction benefits from physicality and the use of human natural motor skills to ease coordination within the interaction space \cite{Cordeil:2017}. One limitation of tangible lenses, though, are their rather fixed size and shape. In terms of lens interaction, these limitations introduce challenges especially when dealing with convoluted volumetric data with cavities, holes, or folds. For instance, it may be difficult to precisely sit a lens into surface cavities or to orient it in areas of high curvature.

To afford more natural interactions, there is a growing interest in incorporating lenses in modern augmented \cite{Brown:2006, Looser:2007} and virtual reality \cite{Tong:2017, Mota:2016} systems. Niebling \cite{Niebling:2017} recently proposed an interaction metaphor using 3D-tracked tablets, which closely resembles cutting planes, for continuous slicing of computational fluid dynamics (CFD) data sets. Advantages of these modern technologies for 3D data exploration include optical tracking (e.g., head and hands) for direct manipulation in 3D; stereoscopy and motion parallax; and proprioception/kinesthesia for body-relative spatial interactions, often in a physical 1:1 scale. Our work leverages the emerging generation of VR in order to minimize interaction effort for lens-based exploration, a major challenge faced by 2D interaction systems.

\setlength\belowcaptionskip{-3ex}

\begin{wrapfigure}{l}{3.3cm}
\centering
\vspace{-0.5cm}
\includegraphics[width=1.55in]{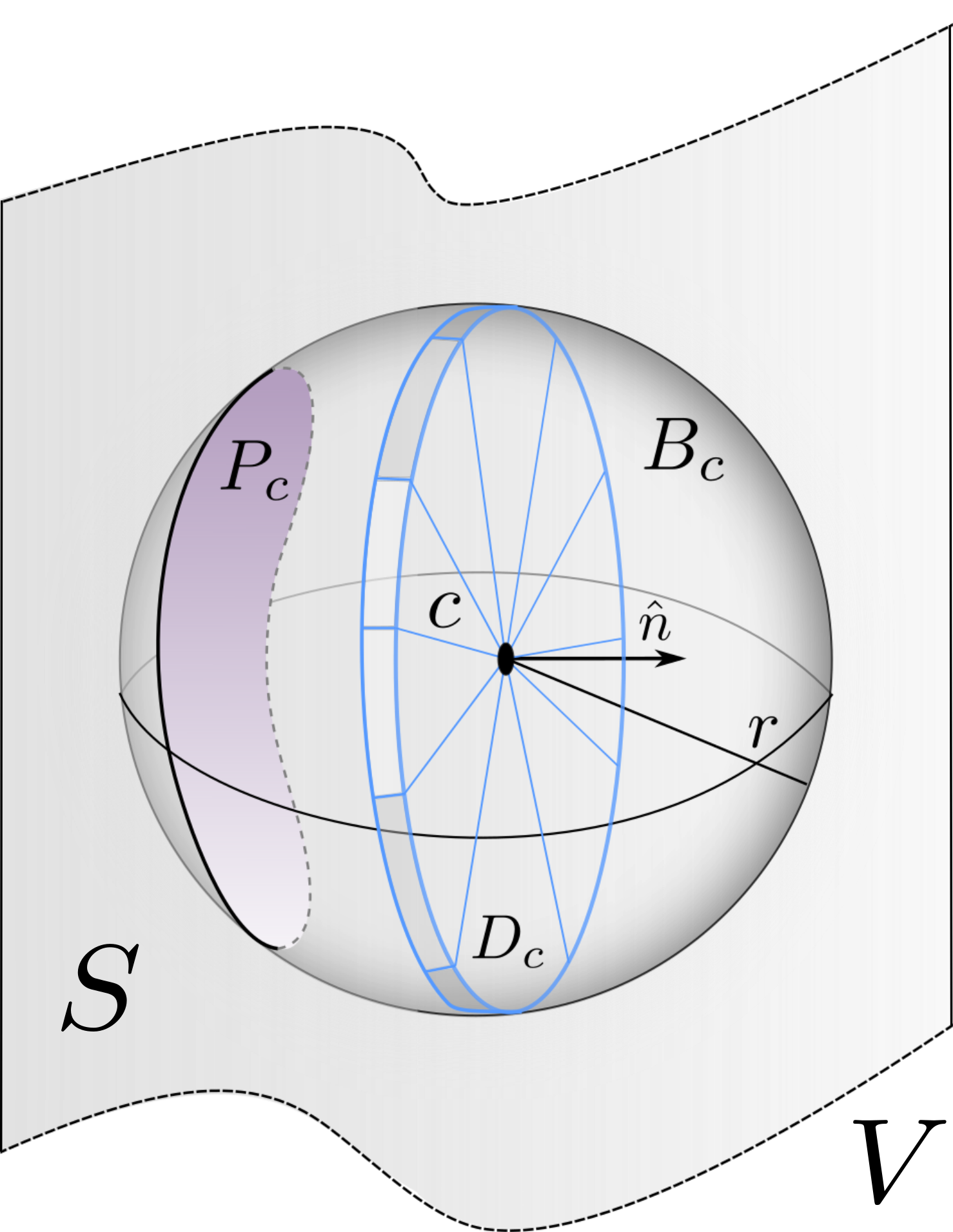}
\vspace{-0.1cm}
\caption{Lens illustration.}
\label{fig:lensConcept}
\vspace{0.2cm}
\end{wrapfigure}

\begin{figure*}[ht!]
\centering 
    \includegraphics[width=6.9in]{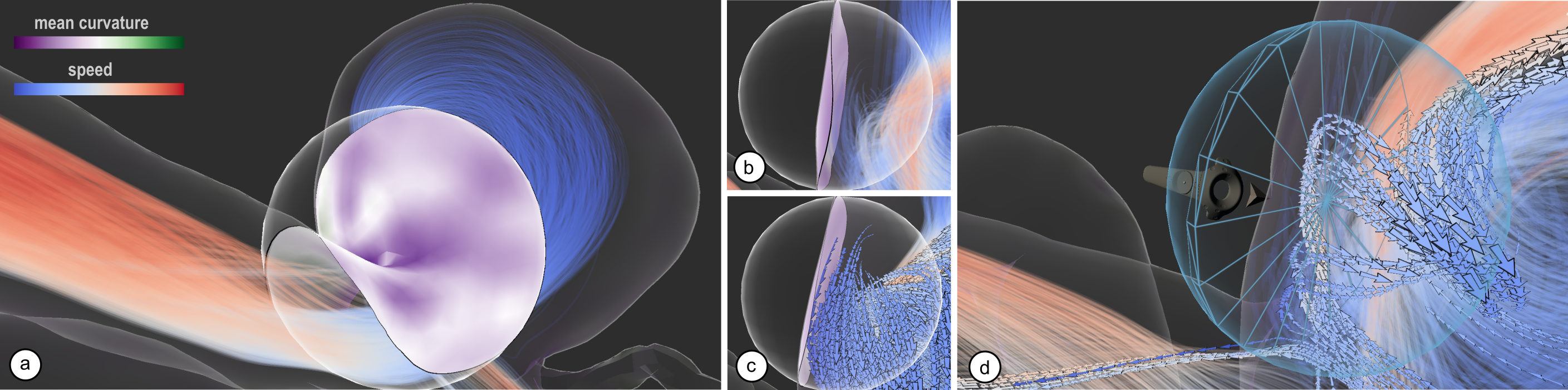}
   \caption{\textit{3De lens} in the aneurysm scope. (a) Decal displaying mean curvature over an aneurysm vessel; the lens patch provides depth cues for perception of surface shape and spatial relations between vessel and underlying blood flow. 
    (b) $L_{De}$ depicts mean curvature on surface focus area. (c) $L_{3D}$ selects near-wall flow described by the magnitude of velocity. (d) $L_{3D}$ further identifies clusters of flow lines following a common direction and focus on a single flow pattern.}
   \label{fig:A1_1}
\end{figure*}


\section{Lens Concept}
Based on the discussion on using different lens categories for 3D data, we propose \textit{3De lens} as a combination of 3D and Decal lenses into a single one. This integration scheme enables the lens to act on different types of data, each of which benefiting from particular traits of its associated lens. 
We refer the reader to Fig. \ref{fig:lensConcept} to illustrate the concepts we now describe.


Let us consider a three-dimensional Euclidean space ${\rm I\!R}^{3}$ containing a volume $V$ and a surface $S$ onto which the 3De lens $L$ will be placed. We denote the lens region as $B_{c}$, a ball centered at an arbitrary point $c$ with radius $r$. The lens $L = \langle L_{3D}, L_{De} \rangle$ is a set of containment lenses:

\begin{itemize}[leftmargin=*]

\setlength{\itemsep}{0pt}
\setlength{\parskip}{0.0pt}

\item $L_{3D}$. $B_{c}$ represents the region that will contain the 3D lens; which allows for selecting a subvolume of $V$ while displaying a non-surface focus attribute $a_{1}$.\\
We also denote a disk within $B_{c}$ as $D_{c}$, with normal vector $\hat{n}$, center $c$, and radius $r$ so that the radial extension of the disk is restricted to the ball shape. $D_{c}$ defines the orientation of the 3D lens; which enables an optional angular selection over $V$ based on the disk's normal direction.

\item $L_{De}$. The intersection of $S$ and $B_{c}$ defines a patch $P_{c} = B_{c} \cap S$. $P_{c}$ can be understood as the area that will contain the decal lens: a 2D submanifold of $S$.
This lens supports the task of selecting a region over $S$ while encoding a surface focus attribute $a_{2}$.

\end{itemize}

\section{Lens Technique}
In this section, we define design goals (DGs) based on the identified limitations of using different interaction modalities and lens techniques for 3D data. Afterwards, we build upon our guidelines to describe the visual and interaction design of our lens technique. To aid our discussion, we use a public aneurysm data set \cite{VisItAneurysm} that contains pressure and mean curvature of the aneurysm vessel as surface attributes; and we trace streamlines to capture the flow profile inside the vessel, with velocity and vorticity as associated attributes. The aneurysm scenario contains two geometric representations: an isosurface and streamlines. It is important to highlight the general nature of our lens technique as it can be applied to other representations (e.g. point clouds and volumes) using the same paradigm.

\begin{itemize}[leftmargin=*]

\setlength{\itemsep}{0pt}
\setlength{\parskip}{0.0pt}

\item\textbf{DG\textsubscript{\textit{Manip.}}} \textit{Facilitate manipulation}. For 3D exploration, to support lens manipulation in a way that minimizes user effort is desirable. Our intent is to use direct 3D manipulation so that interaction operations are seamlessly embodied in the 3D visualization.

\item \textbf{DG\textsubscript{\textit{Fluid.}}} \textit{Provide for fluid interaction}. Interaction is the catalyst for the interplay between data and user, and is an essential factor for lens-based data exploration. We strive for fluid interaction by  \textit{minimizing indirection in the interface} and \textit{providing immediate visual feedback on interaction}, as proposed by Elmqvist \textit{et al.}~\cite{Elmqvist:2011}.

\item \textbf{DG\textsubscript{\textit{Occl.}}} \textit{Minimize occlusion as necessary}. Clutter and occlusion are major challenges in 3D visualization. We thus focus on providing visualizations with minimal clutter in a way that does not compromise depth cues required to discriminate 3D saliency and relationships.

\end{itemize}


\subsection{Interaction Design}
The potential for direct spatial manipulation led us to incorporate our lens technique in VR, using Unity \cite{Unity3D} and an HTC Vive \cite{HTCVive}. Its room-scale tracking allows users to walk around or inside the data, and seamlessly interact with it using tracked headset and controllers.

From an interface perspective, our lens technique captures input/output events from different lenses, keeps track of which lens produced which event, and places all events on a single visual entity. This prevents manipulating multiple independent widgets that compete for screen space, and decreases context switching as user's attention can remain focused on the work area (\textbf{DG\textsubscript{\textit{Fluid.}}}).



We now return our attention to the focus+context visualization of the aneurysm scenario in order to describe our lens interaction vocabulary in VR. First, the user can press and hold the trigger button on the controller to \textit{grab} the lens $L$ so that its center $c$ mirrors the controller's spatial position; and this translates to the constituent lenses since $L$ employs an enclosing structure. The grabbing metaphor supports direct lens manipulation in 3D (\textbf{DG\textsubscript{\textit{Manip.}}}), and is known to be very easily understood due to its natural ``kinaesthetic correspondence'': an isomorphic correspondence between hand movement and the visually perceived motion \cite{Ware:1988, Poupyrev:2001}. While grabbing $L$, the user can seek for surface areas of interest: whenever the lens intersects a surface geometry, $L_{De}$ depicts the surface focus attribute. Fig. \ref{fig:A1_1}-a illustrates how a decal-lens adapts to the surface geometry even in areas of high curvature; and displays  the mean curvature using a diverging purple-to-green colormap.

The user releases the button to \textit{ungrab} $L$. This action triggers events via $L_{3D}$: the selection of all streamlines passing through its lens region and the display of non-surface focus attribute. Fig. \ref{fig:A1_1} shows flow velocity magnitude using a diverging cool-to-warm colormap; we use animated arrow glyphs to communicate the direction of the selected flow near the vessel wall.

Whenever the user places the controller inside $L$, a disk-like widget appears that orients $L_{3D}$. By pressing and holding the trigger button, the user \textit{grabs} the disk so that its normal direction mirrors the controller's upward orientation (\textbf{DG\textsubscript{\textit{Manip.}}}). This in turn produces an angular-based selection carried out via $L_{3D}$: only streamlines that are approximately aligned with the normal of the disk will be selected (see Fig. \ref{fig:A1_1}-d). The angular tolerance allowed is 15 degrees.

Lastly, one can \textit{scale} the lens $L$ in order to alter its radius $r$; this transformation transfers to all enclosed lenses. This is also useful for keeping costs at a level that facilitates interactivity and comprehensibility at all times. This concerns computational costs (\textbf{DG\textsubscript{\textit{Fluid.}}})  but also cognitive costs, i.e., the effort required to make sense of the lens effect (\textbf{DG\textsubscript{\textit{Occl.}}}). In conjunction with the disk-like widget, a small ball shape appears at the lens' center $c$ whenever the user places the controller inside $L$. By grabbing the ball widget and moving their arms to right/left, the user scales up/down $L$.

\subsection{Visual Design and Implementation}
\begin{figure}[t!]
\centering
   \includegraphics[width=\columnwidth]{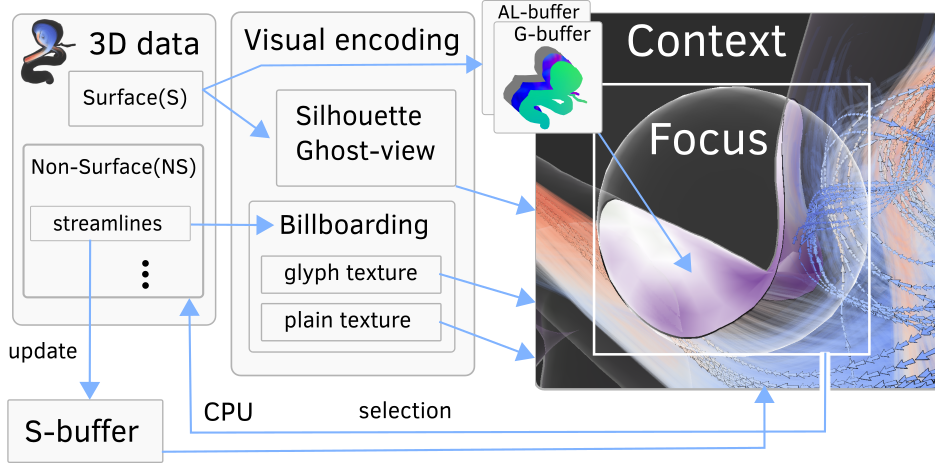}
   \caption{Visualization pipeline}
   \label{fig:overview}
\end{figure}

The HTC Vive uses a display resolution of 2160x1200 (1080x1200 per eye) at a 90 $\mathrm{Hz}$ refresh rate. The combination of high-resolution and frame-rate requirements for VR presents a significant challenge compared to traditional desktop visualization, where low frame rates and intermittent pauses for computations or data loading are more tolerable.  We, therefore, rely on  GPU features such as rasterization functionality, fragment and stencil tests, and blending operations (\textbf{DG\textsubscript{\textit{Fluid.}}}). We obtain interactive frame rates (45 FPS) using a desktop with an Intel\textregistered Core~\texttrademark i3 processor with a GeForce GTX Titan 6G GPU. Our GPU-based implementation is explained via the aneurysm data; this contains 15K surface triangles and 2K blood flow streamlines. We refer the reader to Fig.~\ref{fig:overview} to illustrate the discussion that follows.

\noindent\paragraph{\textit{Context.}}In our visualization, context consists of one or more semi-transparent regions to lessen occlusion while revealing inner, opaque focus regions (\textbf{DG\textsubscript{\textit{Occl.}}}). Concomitantly, important characteristic parts of the context (e.g., contours) are visualized to accommodate better understanding of spatial relationships between structures in the data.

In our implementation, we first read all data (attribute and polygonal geometry) from a volume data set and semantically classify them either as \textit{surface} (S) or \textit{non-surface} (NS) data (Fig.~\ref{fig:overview}-left). These steps happen in a pre-processing stage before the actual rendering process. In the aneurysm scenario, (S) is the vessel surface and (NS) refers to the blood flow streamlines.

We employ multi-pass techniques to render surface information (Fig.~\ref{fig:overview}-middle). The surface outline is first drawn using a two-pass silhouette rendering algorithm similar to Rustagi~\cite{Rustagi:1989}. The first pass enables the stencil buffer and renders the surface; this creates a binary mask of the surface and the background. In the second pass, we extrude the surface vertices along the normal direction (in camera space) in the geometry shader. The stencil buffer visibility test is used to mask out surface fragments, and the remaining fragments compose the surface silhouette. We then draw the surface using a Fresnel-reflection model that maps reflection to opacity values, similarly to Gasteiger \textit{et al.}~\cite{Gasteiger:2010}. In this approach, the Fresnel opacity $F_{o}$ assigned to the surface  is given by $F_{o} = 1 - |\hat{v}\ .\ \hat{n}|\ ^{r}$, where $\hat{n}$ is the surface normal, $\hat{v}$ is the view vector, and $r >= 0$ is the edge fall-off parameter; we obtained good results using $r$ values around $0.5$. The combination of silhouette and Fresnel shading conceive an illustrative representation of the surface (Fig.~\ref{fig:A1_1},~\ref{fig:overview}); these techniques together minimize occlusion by achieving more opacity in regions facing away from the viewer and more visibility in regions facing towards the viewer (\textbf{DG\textsubscript{\textit{Occl}}}).

In the aneurysm scenario, non-surface information consists of streamlines. Similarly to Schirski \textit{et al.}~\cite{Schirski:2004}, we employ a piecewise view-aligned billboarding technique to increase rendering speed and allow for interactive visualization of a large number of lines (\textbf{DG\textsubscript{\textit{Fluid.}}}). The billboards are quadrilaterals computed on the GPU. For each point $p$ along a streamline, two vertices are sent to the graphics system. The upper and lower vertices, $p_{0}$ and $p_{1}$, related to the original position $p$ are determined as $p_{i} = p + r(-1)^{i}(\hat{v}\ \times\ \hat{d})$, ($i \in \{0, 1\}$); where $r$ is the line thickness, $\hat{v}$ is the vector to the viewer, and $\hat{d}$ is the directional vector from the current to the next point on the streamline. Billboard textures are then set based on a selection discrimination: a semi-transparent plain texture for unselected data; and an opaque, animated arrow glyph to convey movement of particles through the flow domain for selected data (Fig.~\ref{fig:overview}-right). A coloration is also applied to encode the non-surface focus attribute. In Fig. \ref{fig:A1_1}, the flow magnitude is represented using a cool-to-warm colormap.



Furthermore, the selection discriminator resides in a uniform buffer object (S-buffer); whose access identifiers are sent to the graphics system as vertex attribute. The number of elements in the buffer depends on the 3D data of input. 
In the aneurysm scenario, we employ a binary selection in which each streamline is either entirely selected or not; the buffer size thus corresponds to the number of lines in the blood flow field. Each point along a streamline stores the unique \texttt{id} of its seed, which is used as a per-vertex attribute on the GPU to access the S-buffer.

\noindent\paragraph{\textit{Focus.}}The lens is designed to convey shape perception while minimizing visual interference (\textbf{DG\textsubscript{\textit{Occl.}}}). The lens is drawn in two steps: we first draw the lens surface using Fresnel shading (with $r = 3$); afterwards, we generate a decal-lens on the surface~(Fig.~\ref{fig:overview}-right). The second step is explained in more detail below.

Similarly to Rocha \textit{et al.}~\cite{decalLenses}, we build two view-dependent buffers consisting of the \textit{surface} data: a geometry buffer (G-buffer) and an attribute-layer buffer (AL-buffer). The G-buffer is a framebuffer object with three attachments: vertices, normals, and depth. The AL-buffer consists of a 2D texture array; the number of 2D textures is equal to the number of surface attribute layers. By applying a multi-pass process (in the draw call), we render each attribute and its visual representation to the AL-buffer. After this off-screen step, all surface attributes are available in screen space as 2D textures.


Afterwards, we can create a decal-lens using both the G-buffer and the AL-buffer. For each pixel within the projected lens' bounding box, we access the corresponding surface position $p$ from the G-buffer and determine whether $p$ lies inside the lens centered at $c$ with radius $r$. The batch of all valid $p$ inside the lens corresponds to the decal-lens region in screen space. We then map the visual representation of the surface attribute using the AL-buffer. In Fig. \ref{fig:A1_1}, the mean surface curvature is displayed using a large decal-lens.

Finally, we render the lens' widgets as wireframes following the approach proposed by Baeretzen \textit{et al.}~\cite{Baerentzen:2006}. This technique renders both the model and its wireframe lines in a single pass inside the rasterization pipeline.



\section{Preliminary Case Analysis}
In this section, we illustrate the use of our \textit{3De lens} to inspect wind turbine aerodynamics using a publicly available data set~\cite{computecanada}; we refer the reader to the supplementary video for complete frames. One's main interest here is to answer whether kinetic energy transported by the wind is efficiently transformed into rotational kinetic energy for each set of blades, e.g., in view of improving the economic viability of renewable energy generation.

The data at our disposal consist of wind turbines (total of 362K surface triangles) and air flow (5K streamlines); pressure is defined on the blades (surface
attribute) and streamlines are computed from fluid speed (volumetric
attribute). By using a 3De lens, we first investigate whether the pressure gradients in the blades
are strong---a requirement for efficient energy conversion.  In Fig.~\ref{fig:airFlow}-left, the leftmost blade has a clear, strong pressure gradient (represented with a
cool-to-warm colormap), which amounts to satisfactory energy transfer to
this blade.  Alas, one can see from the same picture that the wind flow
transitions from laminar to possibly turbulent near the middle of the turbine
(centre of the figure); this may reduce the turbine efficiency as
disorderly flow is unlikely to create the required strong pressure gradients.
Further inspection with the 3De lens confirms small pressure gradients on
the other blades, thus suggesting an unsatisfactory energy conversion. By inspecting how the fluid flow behaves past the first blade with the lens (by filtering streamlines passing through the lens and retaining only those with the same direction as the lens orientation), we observe that such disorderly behaviour is due to the left and centre blades interacting with the flow (Fig. \ref{fig:airFlow}-right).

\section{Conclusion and Future Work}
We presented \textit{3De lens}, a focus+context technique for multi-geometry 3D visualizations. It combines 3D and Decal lenses, and allows flexible use of either one or both lenses on data types they best operate on. We demonstrated the potential use of our lens on two domain visualizations, with surfaces and streamlines. For future work, we would like to tackle other domain scenarios and data geometries. We also plan to support the use of simultaneous multiple lenses, investigating ways of easily applying and composing 3De lenses. In this sense, we are particularly inspired by Rocha \textit{et al.}'s composite operations for decal-lenses~\cite{decalLenses}---e.g., one could brush or lasso over the surface to conceive lens-regions of arbitrary shapes.

Moreover, we discussed the major problem faced by 2D interaction systems of manipulating lenses in 3D, and how it can be aided by VR technology. We plan to expand upon 3D interactions that allow users to work through or parametrize the lens. For instance, to enable one to set a local parametrization on the decal patch as to condition an angular-based selection over a surface; or even to  adjust the angular tolerance to select wider/narrower bundles of curves. \cite{SumrowArticle, SumrowJournal, ntva, VisualAnalytics, VirtualReality, canadaEconomy}


\bibliographystyle{abbrv-doi}

\bibliography{template}
\end{document}